\documentclass[twocolumn,american,aps,prl,superscriptaddress,floatfix]{revtex4-1}

\usepackage{amsmath}
\usepackage{amssymb}
\usepackage{amsthm}
\usepackage{ulem}
\usepackage{bm}
\usepackage{subfigure}
\usepackage[hypertex]{hyperref}
\usepackage{graphicx}
\usepackage{xcolor}
\usepackage{comment}
\usepackage{epstopdf}
\begin{document}

\title{Symmetry Protected Topological Order by Folding a One-Dimensional Spin-$1/2$ Chain}

\author{Pejman Jouzdani}

\affiliation{Department of Physics, University of Central Florida,
  Orlando, Florida 32816, USA}

\date{\today}

\pacs{03.67.Lx, 03.67.Pp, 03.65.Yz, 05.50.+q}

\begin{abstract}
We present a toy model with a Hamiltonian $H^{(2)}_T$ on a folded one-dimensional spin chain. 
The non-trivial ground states of $H^{(2)}_T$ are separated by a gap from the excited states. 
By analyzing the symmetries in the model, we find
 that the topological order is protected
by a $\mathbb{Z}_2$ global symmetry. 
However, by using perturbation series and excluding thermal effects, we show that the $\mathbb{Z}_2$ symmetry  is 
stable in comparison to a standard nearest-neighbor Ising model with a Hamiltonian $H_I$.  
We find that $H^{(2)}_T$ is a member of a family of Hamiltonians that are adiabatically connected to $H_I$.
Furthermore, the generalizations of this class of Hamiltonians, their adiabatic connection to $H_I$, and the relation 
to quantum error-correcting codes are discussed. 
Finally, we show the correspondence between the two ground states of $H^{(2)}_T$ and the unpaired Majorana modes, and provide 
numerical examples.
\end{abstract}
\maketitle

\emph{Intro.}--
A large effort has been made to construct robust protocols for
quantum computation by exploiting topological properties of many-body
systems \cite{Sarma, KitaevA}. Models such 
as the toric and surface codes \cite{Dennis, kitaevT, bravyi1998} are proposed. 
A particular attention has been
given to detect and employ exotic non-Abelian excitations in
quantum computation \cite{Nayak}, with a special attention to the  Majorana
fermions \cite{Jason, Fu, ginossar, Buhler, Kraus}. 
In this order, a crucial step is to detect and characterize the topological orders. 
It has been shown by Levin and Wen \cite{LevinWen} that the ground state
of quantum many-body systems with non-trivial orders 
can be seen as a condensate of fluctuating string-like objects. 
In particular, quantum phases are commonly studied by the projective 
symmetry groups (PSG) tool \cite{PSGWen} which has been used to identify 
symmetry protected topological orders (SPT)  \cite{Xie, Afflek, Chen, Son, Pollmann}, and with 
focus to quantum computation and quantum error correction \cite{ Akimasa, Else}.
  
Consider a one-dimensional spin-$1/2$ Ising model with the Hamiltonian
\begin{eqnarray}
\label{eq:IsingH}
H_I = -J \sum_{i=1}^{N-1} \sigma^z_i \sigma^z_{i+1},
\end{eqnarray}
on a chain with 
$N$ spins ($J>0$) where $\sigma^\nu_i$ is  the $\nu$-th component of the Pauli matrices acting on site $i$.
Ignoring thermal excitation for a moment, 
a longitudinal field perturbation $U_z=\sum h_i \sigma^z_i$ lifts the
degeneracy of the ground states for any non-zero $h_i$. However, in the absence of a longitudinal field, the degeneracy is topologically protected 
aginst a perturbation  
$U_x=\sum h_i \sigma^x_i$. 

In comparison to the toric code \cite{kitaevT} where the system is defined on a $L\times L$ lattice and 
the protection against perturbation is of the order of $L$, 
we could think of the one-dimensional Ising chain as 
a $1\times N$ two-dimensional lattice. The lattice has a width of only ``one'' lattice site. Therefore, 
any (longitudinal) single-spin perturbation already reaches the size of the lattice.
Thus, the Ising chain has a topological phase, but the phase can not be realized due 
to the short width of the lattice. Although in different words, this claim was originally 
expressed in a footnote in Ref. \cite{Yu}.  
        
In this Letter we introduce an adiabatic transformation that does not change the topological characteristic of 
the Ising Hamiltonian $H_I$, but effectively folds the spin chain to a $2\times \frac{N}{2}$ lattice. 
As a result, we obtain a Hamiltonian $H^{(2)}_T$ and the topological 
protection of the one-dimensional Ising chain tremendously improves, unexpectedly. 
In order to do so, the steps bellow are followed. 

First, a Hamiltonian $H^{(2)}_T$ on an open chain 
of spin-$1/2$ is defined. $H^{(2)}_T$ is a sum of four-spin operators. The set of these operators form a group that is denoted by ${\cal S}^{(2)}$. 
In addition, we find two symmetry groups ${\cal S}_1^{(2)}$ and ${\cal S}_2^{(2)}$ 
that commute with all the elements of  ${\cal S}^{(2)}$.
We find that  $({\cal S}^{(2)}_1 \otimes {\cal S}^{(2)}_2 ) / {\cal S}^{(2)}= \mathbb{Z}_2$ 
which indicates the non-trivial order of the model is protected as long as 
the global symmetry $\mathbb{Z}_2$ is preserved. 
Next, we show through a degenerate perturbation analysis that the global symmetry $\mathbb{Z}_2$ is robust. Then, 
the extension to models with wider width $H^{(width)}_T$ and with symmetry groups $\{S^{(width)}_i\}$ is discussed.
Finally, we show the Hamiltonian $H^{(2)}_T$ is adiabatically connected to $H_I$ in Eq. (\ref{eq:IsingH}),
\begin{eqnarray}
\label{eq:unitarytransf}
H^{(2)}_T = R(\pi)\,H_I\,R^\dagger(\pi),
\end{eqnarray}
where $R(\alpha)$ is the unitary transformation
\begin{eqnarray}
\label{eq:unitarytransf2}
R(\alpha)=e^{i \frac{\alpha}{4} V}.
\end{eqnarray}
Here $V$ is a sum of two-spin interactions and $\alpha$ is a scalar parameter.
The relation between Majorana modes of the Kitaev toy model \cite{Yu} and the two ground
states of $H^{(2)}_T$ are explained as well.  

\emph{The Model.}-- 
\begin{figure}[h]
\centering
\includegraphics[width=8cm, height= 5cm]{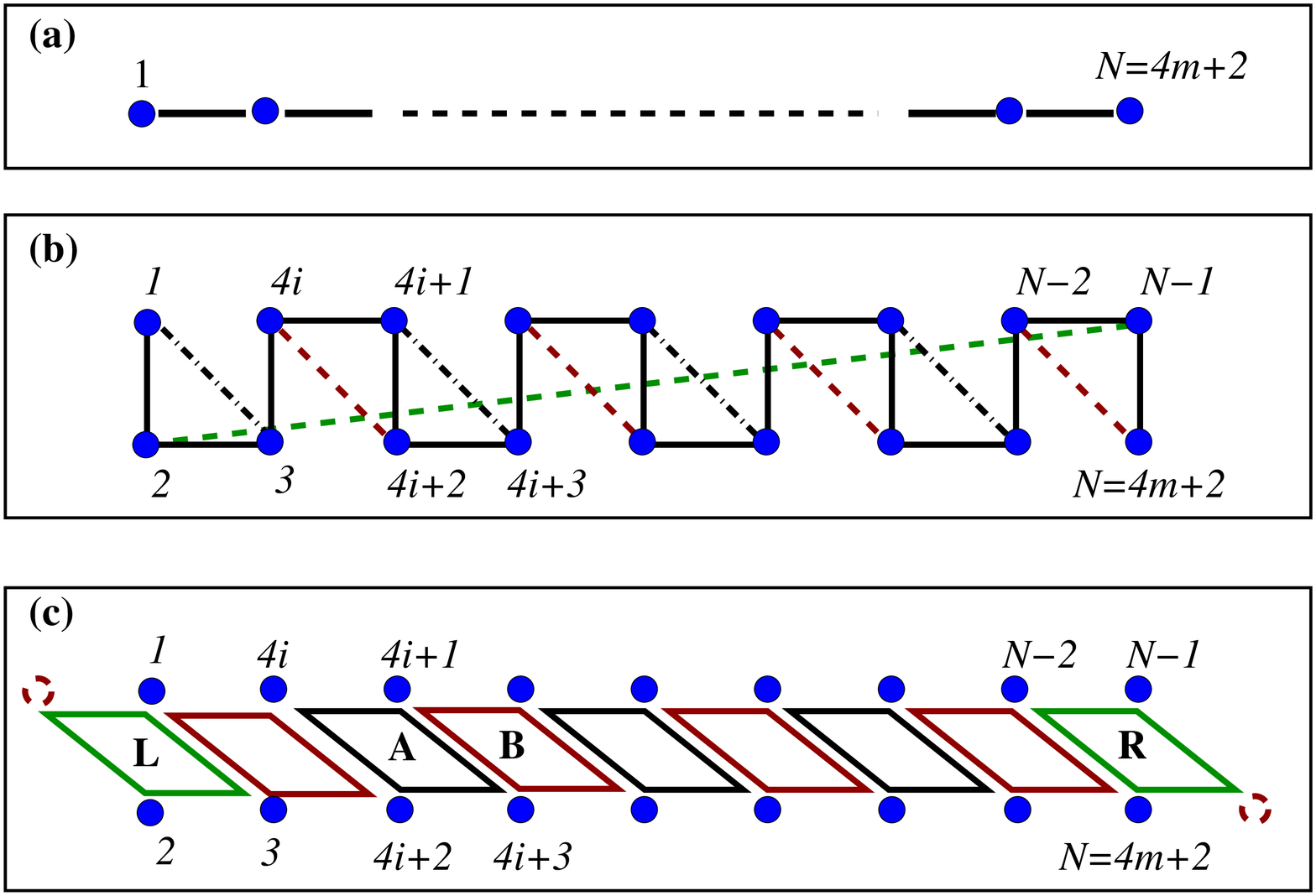}
\caption{(a) The standard one-dimensional Ising model with
  nearest-neighbor interactions. (b) The reshaped Ising model into a
  ladder after the unitary transformation $R$ in spin space. The
  diagonal dashed and dash-dot lines show the two-spin terms
  $\sigma^y_i\sigma^y_{j}$ used in the transformation. (c) The Hamiltonian $H^{(2)}_T$
  resulting from the transformation has four-spin
  interactions, associated to plaquettes. The bulk has two different
  types of plaquettes, $A$ and $B$. There are three operators of type
  $A$ and one operator of type $B$. The two left and right plaquettes
  on the boundaries are named $R$ and $L$, respectively. There are two
  operators acting on each of these boundary plaquettes.}
\label{fig:QNA} 
\end{figure}
Consider a one-dimensional spin-1/2 system of
a length $N=4m + 2$, with a positive and integer number $m$. We define $H^{(2)}_T$ as
\begin{eqnarray}
\label{eq:GHT}
H^{(2)}_T &=& \sum_s O_s \nonumber \\
			 &=&-J \sum_{i=1}^{m-1} \left[
  A_{i}^{(1)} +A_{i}^{(2)} + A_{i}^{(3)} \right] - J \sum_{i=0}^{m-1}
B_{i} \nonumber \\ & & -\ J\, \left( L_1 + R_1 + L_2 + R_2 \right),
\end{eqnarray}
where we have introduced the following operators (the stabilizers $\{O_s\}$):
%
\begin{subequations}
\label{eq:plaquetteR2}
\begin{align}
A_{i}^{(1)} & =  \sigma^x_{4i}\sigma^x_{4i+1}\sigma^y_{4i+2} \sigma^y_{4i+3} \\
A_{i}^{(2)} & =  \sigma^y_{4i}\sigma^y_{4i+1}\sigma^x_{4i+2} \sigma^x_{4i+3}   \\
A_{i}^{(3)} & =   \sigma^y_{4i}\sigma^x_{4i+1}\sigma^x_{4i+2} \sigma^y_{4i+3}   \\
B_{j} & =  \sigma^y_{4j+1}\sigma^x_{4j+3}\sigma^x_{4j+4} \sigma^y_{4j+6} \\
L_1 & =  \sigma^x_{1}\sigma^x_{2} \sigma^y_{3} \sigma^y_{N-1}  \\
R_1 & =  \sigma^y_{2}\sigma^y_{N-2}\sigma^x_{N-1} \sigma^x_{N}  \\
L_2 & =  \sigma^y_{1}\sigma^x_{2}\sigma^x_{3} \sigma^y_{N-1}  \\
R_2 & =  \sigma^y_{2}\sigma^x_{N-2}\sigma^x_{N-1} \sigma^y_{N},
\end{align}
\end{subequations}
with $i=1,\ldots,m-1$ and $j=0,\ldots,m-1$. 

All the terms on the r.h.s. of Eq. (\ref{eq:GHT}) commute with each other. 
As shown in Fig. \ref{fig:QNA}c, one can identify $A_i^{(k)}$ and $B_i$ as bulk plaquettes operators, while
$R_{1,2}$ and $L_{1,2}$ act as boundary plaquette operators. 
For $N=4m+2$ there are $3(m-1)$ plaquette operators of type $A$, $m$ plaquette operators of
type $B$, and two plaquette operators of types $R$ and $L$ each. 
Therfore, there are overall $4m+1$ plaquette operators - that is $N-1$. 
Without proof, the set of the stabilizers  
defined in Eq. (\ref{eq:plaquetteR2}) generates a group that we denote by  ${\cal S}^{(2)}$.

Separately, consider the set of the operators $ \{ \sigma^z_i \sigma^z_{j} \}$, on pairs $\{(i , j)\}$.  
Unless otherwise mentioned, we use the notation $(i,j)$ for the two sites $i$ and $j$
that are connected with a dashed or a dot-dash line in Fig. \ref{fig:QNA}b.
We define the group generated by these operators as
${\cal S}_1^{(2)}= \langle \sigma^z_1\sigma^z_3, \dots, \sigma^z_{N-2}\sigma^z_{N}, \sigma^z_2\sigma^z_{N-1}\rangle$.
All elements in ${\cal S}_1^{(2)}$ commute with all elements in ${\cal S}^{(2)}$. 

Furthermore, consider the set of operators $ \{\sigma^x_i \sigma^y_{j} \}$ on pairs $\{(i , j)\}$. 
We define the group generated by these operators as
${\cal S}^{(2) }_2= \langle \sigma^x_1\sigma^y_3, \dots, \sigma^x_{N-2}\sigma^y_{N}, \sigma^x_2\sigma^y_{N-1}\rangle$.
All elements in ${\cal S}^{(2) }_2$ commute with all elements in ${\cal S}^{(2) }$ and all elements in ${\cal S}^{(2) }_1$. 
Especially, we find  $({\cal S}^{(2)}_1 \otimes {\cal S}^{(2)}_2 ) / {\cal S}^{(2)}= \mathbb{Z}_2$
(there are $N/2$ generators in ${\cal S}^{(2) }_2$,
$N/2$ generators in ${\cal S}^{(2) }_1$, and $N-1$ generators in ${\cal S}^{(2) }$; $2^N/2^{N-1}=2$).
A comprehensive classification of SPT orders in one-dimensional spin systems is given in Ref. \cite{Xie}. 

The physical implication of the above statements is the following.
If we manage to have a fixed ``gauge'' ${\cal S}^{(2) }_1|G\rangle =+|G\rangle$ 
for the ground state subspace $|G\rangle$ of 
$H^{(2)}_T$, as long as the $\mathbb{Z}_2$ symmetry is not broken, 
the ground state is doubly degenerate and we can have $|G_{\pm}\rangle$  
such that $\gamma |G_{+}\rangle = e^{i\alpha/2} |G_{-}\rangle$ for any generator $\gamma$ of ${\cal S}^{(2) }_2$ \cite{note1}. The phase 
$e^{i\alpha/2}$ is a global phase and it is equal to $e^{i\pi/2}$ for $H^{(2)}_T$ defined in Eq. (\ref{eq:GHT}).

To see this, consider  the states $\left| \bar{\uparrow}\right\rangle_{\bf z}= \left|\uparrow\right\rangle_1 \otimes
\cdots \otimes \left|\uparrow\right\rangle_N$  
and $\left|\bar{\downarrow}\right\rangle_{\bf z} = \left|\downarrow\right\rangle_1 \otimes
\cdots \otimes \left|\downarrow\right\rangle_N$ where by definition 	
$\sigma^z_i\left|\uparrow\right\rangle_i = +\left|\uparrow\right\rangle_i$ 
and $\sigma^z_i\left|\downarrow\right\rangle_i = -\left|\downarrow\right\rangle_i$. 
Next,  for every element $s_k\in {\cal S}^{(2)}_{1}$ 
we have $s_k \left|\bar{\uparrow}\right\rangle_{\bf z} =+ \left|\bar{\uparrow}\right\rangle_{\bf z}$ 
($s_k \left|\bar{\downarrow}\right\rangle_{\bf z}= +\left|\bar{\downarrow}\right\rangle_{\bf z}$), 
and thus the gauge ${\cal S}^{(2)}_1$ is set to +1. 
Then, by applying the group elements of ${\cal S}^{(2)}$  on each of the states $\left| \bar{\uparrow}\right\rangle_{\bf z}$ and
$\left| \bar{\downarrow}\right\rangle_{\bf z}$ we obtain
\begin{eqnarray}
\label{eq:cond+}
|G_{+}\rangle = \frac{1}{\sqrt{ 4^{m} } }\prod_{s} \left[ 1 + O_s\right] \left|\bar{\uparrow}\right\rangle_{\bf z} 
\end{eqnarray}
and 
\begin{eqnarray}
\label{eq:cond-}
|G_{-}\rangle = \frac{1}{\sqrt{ 4^{m} }} \prod_{s} \left[ 1 + O_s\right] \left|\bar{\downarrow}\right\rangle_{\bf z},
\end{eqnarray}
where the product is over the stabilizers $O_s$. Notice that $2m=\frac{N}{2}-1$ is the number of 
distinguishable plaquettes in Fig. \ref{fig:QNA}c. Thus, $ 4^{m}$
is the number of ``loops'' that can be constructed on the folded chain using the plaquettes as the unit blocks. 
The basis states appear as condensates of string-like configurations \cite{LevinWen}.
Interestingly, for a generator $\gamma=\sigma_i^x\sigma^y_j$ of ${\cal S}^{(2) }_2$
\begin{eqnarray}
\label{eq:globalphase}
\gamma |G_{+}\rangle =  i |G_{-}\rangle,
\end{eqnarray}
since $\sigma^y|\uparrow\rangle=+i |\downarrow\rangle$.  The equations (\ref{eq:cond+}), (\ref{eq:cond-}), 
and the property in Eq. (\ref{eq:globalphase})
can be examined in the examples given at the end of this Letter. 

In fact, the property in Eq. (\ref{eq:globalphase}) is not a coincidence if one 
remembers that by a Jordan-Wigner transformation the Ising model maps to the unpaired Majorana problem \cite{Yu}. Thus, the generators in the group 
 ${\cal S}^{(2) }_2$ act equivalently as a logical operation for the ground states of $H^{(2)}_T$.

\emph{Stability of the $\mathbb{Z}_2$ symmetry}.-- In a one-dimensional  Ising spin on an open chain a non-zero longitudinal field
opens a gap between $|\bar{\uparrow}\rangle$ and $|\bar{\downarrow}\rangle$. The gap stimulates 
topological excitations (a propagating domain wall) from a \emph{false vacum} to a \emph{true} vacum
and eventually destroys the symmetry \cite{Simons}. 

Similar to $H_I$, the Hamiltonian $H^{(2)}_T$ has a discrete energy spectrum and at low temperature ($k_BT\ll J$) 
thermal excitations are energetically costly 
and can be considered forbidden.
By a degenerate perturbation series approach we see that a transverse 
field perturbation such as  $U_{x}=\sum_i h_i \sigma^{x}_i$ (or $U_{y}=\sum_i h_i \sigma^{y}_i$) 
has vanishing matrix elements 
$\langle G| U^l_{x}|G\rangle$ for all the powers $l<N$, in the ground states supspace 
($|G\rangle \equiv |G_+\rangle \langle G_+ | + |G_-\rangle\langle G_-|$). 
Therefore, we have a topological protection with respect to this transversal field. 
Especially, the only non-vanishing term is  $\langle G| X=\prod_{i=1}^N \sigma^x_i |G\rangle$.

Considering the bases $|G_{+}\rangle \pm i \,|G_{-}\rangle$, we find $X=\prod_{i=1}^N \sigma^x_i$ and any generator $\gamma \in {\cal S}^{(2) }_2$ 
as the  \emph{bit-flip} and \emph{phase-flip} logical operations, respectively.

In contrast to $H_I$, in the presence of two transversal fields ($H_T^{(2)} + U_{x} + U_{y}$), the first non-vanishing term 
$\langle G| U_{x}U_{y}|G\rangle \propto \langle G| \sigma_i^x\sigma_j^y |G\rangle $ appears
at the \emph{second order} of the perturbation series and only on pairs $(i , j)$. 
This means that, out of $ N \choose 2$ number of 
possible terms in the second order, only $\frac{N}{2}$ of them are non-zero. 
Thus, by increasing the length, the second order non-vanishing terms are 
suppressed by a factor of ${\cal O} (\frac{1}{N})$.  This is opposite to the 
$H_I + U_z$ case where in the second order of perturbation there are $ {N \choose 2} - (N-1)$
non-vanishing terms (${\cal O} (N)$). This pattern continues in all the orders. The odd 
orders vanish. In the fourth order
there is a suppression factor of ${\cal O} (\frac{1}{N^2})$, etc. 
  
Thus, as long as the perturbation (noise) affects single spins (no correlated noise) 
we should expect that the interplay between multiplicity and energy cost in the perturbation series (the statistical ground for a phase transition)
to be substantially suppressed in our model with $H^{(2)}_T$ Hamiltonian. 
Therefore, we expect a stable global  symmetry $\mathbb{Z}_2$.
Whether the $\mathbb{Z}_2$ symmetry will still be destroyed through other mechanisms
is a question to be answered. For a ladder of the toric code this has been
studied \cite{Karimpour}.

\emph{Translational symmetry breaking and generalization}.--
By a close look at Fig. \ref{fig:QNA}a and Fig. \ref{fig:QNA}c, we notice that the periodicity changes as we
move from the Ising chain in Fig. \ref{fig:QNA}a to $H_T^{(2)}$ in Fig. \ref{fig:QNA}c. The unit cell in $H^{(2)}_T$ is the two-plaques $AB$ and the 
periodicity goes as $\cdots ABAB \cdots$ in the bulk, while in the Ising chain it goes as $\cdots ZZZZ \cdots$. The emergence of 
non-trivial topological orders in one dimensional ``organic'' polymers by breaking translational symmetry has an old history \cite{Su}. 

Also, notice that the width of the folded chain in Fig. \ref{fig:QNA}c contains two sites. It should be now clear to the reader
why we chose $N=4\times m +2$. There are $4$ sites ($2\times width$) 
in each unit cell and the last $2$ sites are added to keep the inversion symmetry and 
have $({\cal S}^{(2)}_1 \otimes {\cal S}^{(2)}_2 ) / {\cal S}^{(2)}= \mathbb{Z}_2$. Although, it is not clear whether the inversion symmetry is 
necessary.

Similarly, we can construct a Hamiltonian $H^{(3)}_T$ with a width of 
``three'' sites and define stabilizers with ``six'' operators \cite{note2}. 
In this case, the length of the chain is chosen to be $N=6m$ with a positive integer $m$. That is, $m-1$ unit cells in the bulk. 
There are five different $A$-type operators, defined on $A$ plaquettes, and one $B$-type stabilizer (in a hexagonal shape), with two boundary 
stabilizers on each edge. Therefore, there are $N-1$ number of stabilizers. 
The non-trivial part is to show that there are two independent symmetry groups $S^{(3)}_1$ 
and $S^{(3)}_2$ that commute with the group of stabilizers of $H_T^{(3)}$, ${\cal S}^{(3)}$. That is, 
they satisfy $({\cal S}^{(3)}_1 \otimes {\cal S}^{(3)}_2 ) / {\cal S}^{(3)}= \mathbb{Z}_2$.

$H^{(3)}_T$ is the first member of the quantum error-correcting codes that can be constructed by folding a line and has a code
distance of ${\cal O} (1)$; That is roughly half of the width ($3$) of the folded configuration.
In principle, it should be possible to construct 
quantum error-correcting codes with longer code distance and larger stabilizers with Hamiltonian $H_T^{(width)}$. As we will see for the case
of $H^{(2)}_T$ bellow, there is an adiabatic connection between this class of Hamiltonians and $H_I$ in Eq. (\ref{eq:IsingH}).


\emph{ The Hamiltonian $H^{(2)}_T$ is adiabatically connected to the one-dimensional Ising Hamiltonian $H_I$}.--       
We begin with the standard nearest-neighbor Ising Hamiltonian in Eq. (\ref{eq:IsingH})
with $J>0$. Next, the unitary transformation $R$ defined in Eq. (\ref{eq:unitarytransf2})  with
\begin{eqnarray}
\label{eq:V}
V & = &  \sum_{(i,j)} 
\sigma^y_{i} \sigma^y_{j} 
\end{eqnarray}
is used  to map $H_I$ to $H^{(2)}_T$ according to Eq. (\ref{eq:unitarytransf}).
The result is
\begin{eqnarray}
\label{eq:Halpha}
H(\alpha) & = & R (\alpha)\, H_I\, R^\dagger(\alpha) \nonumber \\ & =
& \cos^2 \left( \frac{\alpha}{2} \right)\, H_I + \cos
\left(\frac{\alpha}{2} \right) \sin \left(\frac{\alpha}{2} \right)\,
H_1 \nonumber \\ & & +\ \sin^2 \left( \frac{\alpha}{2} \right)\, H^{(2)}_T,
\end{eqnarray}
where $H_1$ and $H^{(2)}_T$ involve three-body and four-body interaction
terms, respectively. For $\alpha=\pi$, the contribution of $H_1$ to the total
Hamiltonian drops out and we obtain $H(\pi)  =  H^{(2)}_T$.

Since we obtain $H^{(2)}_T$ by a  unitary 
transformation from  $H_I$, the energy  spectrum must stay unchanged and the ground state subspace
of $H^{(2)}_T$ must be two-fold degenerate. However, if two gapped states are connected by a set of local 
unitary transformations they belong to the same phase \cite{Chen2}.
This can be seen by applying $R(\pi)$ on a state $| \bar{\uparrow}\rangle$. One obtains 
\begin{eqnarray}
\label{eq:psi}
\left| \psi \right\rangle =R(\pi) | \bar{\uparrow}\rangle = \prod_{(i,j)} 
(\left| \uparrow \right\rangle_i \left|\uparrow\right\rangle_j -\, i \left|\downarrow \right\rangle_i \left|\downarrow \right\rangle_j),
\end{eqnarray}
which is a product state. 
As it can be checked, 
clearly a second degenerate state is not accessible  
by using Eq. (\ref{eq:globalphase}). That is, for a generator $\gamma$ of the group 
${\cal S}^{(2)}_2$, we have $\gamma \left| \psi \right\rangle = -\left| \psi \right\rangle$ while $N= 4m +2$.
This means 
$\left| \psi \right\rangle  = \frac{1}{\sqrt{2}}\left(|G_{+}\rangle - i \,|G_{-}\rangle\right)$.
This is not surprising if one thinks of the Majorana counterpart of the problem where 
the two unpaired Majorana modes ($|G_{+}\rangle$ and $|G_{-}\rangle$) are paired up and  
experimentally undetectable.

Then, how can we observe the phase that corresponds to $|G_{+}\rangle$ (or $|G_{-}\rangle$)? One quick answer is to
find a way to implement logical quantum gates.  
We are seeking an operation such that $M^{(\pm)}\left| \psi \right\rangle = |G_{\pm}\rangle$. 
Since $\gamma$ is a logical phase-flip operation and $X=\prod_{i=1}^N \sigma^x_i$ is the logical bit-flip operation, we should be able to
decompose $M^{(-)}$ and have
\begin{eqnarray}
\label{eq:M}
M^{(-)}=\frac{1}{\sqrt{2}}\left[ \gamma X -i\gamma  \right].
\end{eqnarray}
Notice that the logical operation $M^{(-)}$ is unitary but non-local. 
Implementing $M^{(-)}$, if ever possible experimentally, would change the quantum phase from a product state
in Eq. (\ref{eq:psi}) to a Majorana mode  $|G_{-}\rangle$. This can be checked for a chain with $N=6$ in the examples bellow. 

Notice that we can obtain $\gamma$ and $X$, corresponding to the logical operations of $H^{(2)}_T$, 
by transforming the $X=\prod_{i=1}^N \sigma^x_i$ and the order parameter $\sigma^z_i$ corresponding to $H_I$ under $R$. In general,
to obtain the logical operations corresponding to $H^{(width)}_T$ 
one needs to know the adiabatic transformation from $H_I$ and the knowledge of the 
symmetry groups of the Hamiltonian becomes irrelevant.


\begin{figure}[ht]
\centering
\includegraphics[width=8cm, height = 5cm]{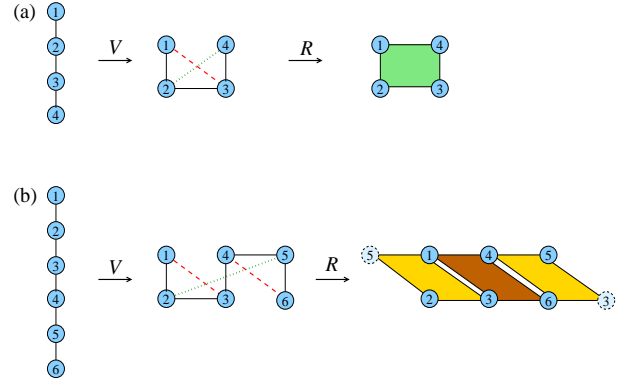}
\caption{Chains with $N=4$ (a) and $N=6$ (b) spins. The solid lines
  represent the initial Ising nearest-neighbor interaction between the
  spins, $H_I$. The dashed and dotted lines indicate the interaction
  terms present in $V$ (Eq. (\ref{eq:V})) which are used to generate the transformation
  $R$.}
\label{fig:physics} 
\end{figure}
%
%
%
%
%

\emph{Examples and numerical results}.-- Let us consider two examples. The first example is a chain of 
$N=6$. Following Eq. (\ref{eq:GHT}), we have (see Fig. \ref{fig:physics}b)
\begin{equation}
\label{eq:hamiltT6}
H^{(2)}_{T,\,N=6} = -J\,(L_1 + L_2 + B_0 + R_1 + R_2).
\end{equation}
By exact diagonalization we obtain the two basis states
\begin{eqnarray}
\label{eq:Psi6}
|G_{+,\, N=6} \rangle &=& 
\frac{1}{\sqrt{4}} [| \uparrow  \uparrow   \uparrow   \uparrow   \uparrow   \uparrow   \rangle  - 
|\downarrow \downarrow \downarrow \uparrow   \downarrow \uparrow   \rangle - 
|\downarrow \uparrow   \downarrow \downarrow \uparrow   \downarrow \rangle \nonumber \\ &-& 
|\uparrow   \downarrow \uparrow   \downarrow \downarrow \downarrow \rangle 
].
\nonumber \\
|G_{-,\, N=6} \rangle &=& 
\frac{1}{\sqrt{4}} [ |\downarrow \downarrow \downarrow \downarrow \downarrow \downarrow \rangle - 
|\uparrow   \uparrow   \uparrow   \downarrow \uparrow   \downarrow \rangle - 
|\uparrow   \downarrow \uparrow   \uparrow   \downarrow \uparrow   \rangle \nonumber \\ &-& 
|\downarrow \uparrow   \downarrow \uparrow   \uparrow   \uparrow   \rangle 
]
\nonumber  \\
\end{eqnarray} 
The second example is a chain with $N=4$. It does not exactly follow the
prescription defined in Eq. (\ref{eq:GHT}). However, we can define  (see
Fig. \ref{fig:physics}a) 
\begin{equation}
\label{eq:hamiltT4}
H^{(2)}_{T,\,N=4} = -J\, (\sigma_1^x\sigma_2^x\sigma_3^y\sigma^y_4
+\sigma_1^y \sigma_2^x\sigma_2^x\sigma^y_4
+\sigma_1^y\sigma_2^y\sigma_3^x\sigma^x_4),
\end{equation}
and the inversion symmetry in this case is still preserved. It has the degenerate basis states 
\begin{eqnarray}
\label{eq:GT+4}
|G_{+,\, N=4} \rangle &=& 
\frac{1}{\sqrt{2}} \left[
| \uparrow \uparrow \uparrow \uparrow \rangle - 
|\downarrow \downarrow \downarrow \downarrow \rangle
\right] 
\nonumber \\
|G_{-,\, N=4}\rangle &=& 
\frac{1}{\sqrt{2}} \left[
|\uparrow \downarrow \uparrow \downarrow \rangle + 
|\downarrow \uparrow \downarrow \uparrow \rangle
\right].
\end{eqnarray}
Notice that by applying $\sigma^x_1\sigma^y_3\in S^{(2)}_2$ on $|G^{(4)}_+ \rangle$ 
($|G^{(6}_+ \rangle$) one obtains $i |G^{(4)}_- \rangle $ ($i |G^{(6}_- \rangle$). 

\begin{figure}
\centering
\includegraphics[width=8cm, height = 5cm]{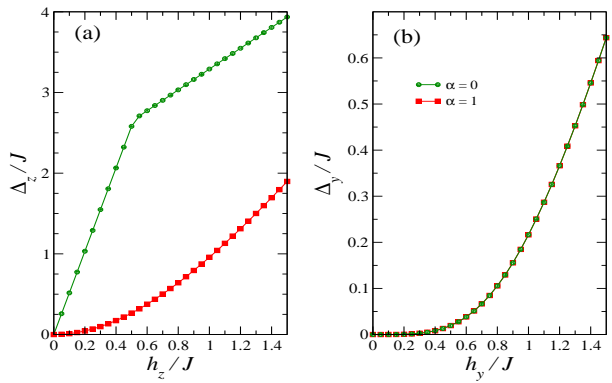}
\caption{The energy  splitting between the two low-lying states at $\alpha=0$
  ($H_I$, circles) and $\alpha =1$ ($H_T$, squares) as a function of
  external field for a $N=4$ spin chain. (a) The splitting as a function of a longitudinal
  field $h_z$ is shown for the two $H_I$ and $H^{(2)}_T$.
  For any non-zero
  longitudinal magnetic field $h_z$, a gap opens linearly for $H_I$.
  The situation is visibly different (quadratic) for
  $H^{(2)}_T$ as discussed by perturbation analysis. 
 (b) The splitting as function of a transverse
  field $h_y$ is the same for both $H_I$ and $H^{(2)}_T$.}
\label{fig:gaps} 
\end{figure}

We use the dependence of the splitting of the ground states,
$\Delta(\alpha) = E_+(\alpha) - E_-(\alpha)$, on the global external
magnetic field as a criterion to numerically verify the enhanced protection in
$H_T$. The numerical calculation of the splitting for $N=4$ at the points $\alpha=0$ ($H_I$) and
$\alpha=\pi$ ($H_T$) is shown in Figs. \ref{fig:gaps}a and
\ref{fig:gaps}b as a function of $h_z$ (longitudinal) and $h_y$
(transverse) external magnetic fields, respectively.   One can see that the dependence
of the splitting $\Delta_z$ goes from linear for $H_I$ to quadratic
for $H_T$, indicating increased protection. Figure \ref{fig:gaps}b
shows the gap $\Delta_y$ which behaves topologically protected for both $H_I$ and
$H_T$, as expected.


\emph{Summary}.-- The standard one-dimensional Ising chain with a Hamiltonian $H_I$ 
could theoretically be in a topological phase if 
the global $\mathbb{Z}_2$ symmetry were stable. 
The symmetry is not stable since the order parameter of the system is just a single spin.
We showed how to adiabatically obtain a Hamiltonian $H^{(2)}_T$ by a local unitary transformation with only two-spin interactions from  $H_I$.
We showed that the
protection against single-spin errors in the transformed Hamiltonian $H^{(2)}_T$ scales with the length of the chain. 
We discussed a family of Hamiltonians that are adiabatically connected to $H_I$. 

\emph{Acknowledgments}.-- The author thanks Eduardo Mucciolo for his advise and support and Alioscia Hamma for his critical comments. 
This work was supported in part by the
National Science Foundation grant CCF-1117241.




\begin{thebibliography}{10}
\bibitem{note1} Notice that ${\cal S}^{(2) }_i \subset {\cal S}^{(2) }$.  Thus, only one generator is enough to represent the whole subgroup.
\bibitem{note2} The details will be published elsewhere.
\end{thebibliography}
\end{document}